\documentclass[11pt]{article}

\usepackage[utf8]{inputenc}
\usepackage[T1]{fontenc}
\usepackage[margin=1in]{geometry}
\usepackage{amsmath,amssymb,amsfonts}
\usepackage{bm}
\usepackage{graphicx}
\usepackage{tikz}
\usetikzlibrary{arrows.meta,positioning,calc}
\usepackage{authblk}
\usepackage[hidelinks]{hyperref}
\usepackage{enumitem}
\usepackage{siunitx}
\usepackage{booktabs}

\graphicspath{{figs/}}

\title{Dense Holographic Associative Memories}
\author{David J. Brady}
\author{Gregory Nero}
\affil{Wyant College of Optical Sciences, University of Arizona, Tucson, Az, USA}

\newcommand{\xivec}{\bm{\xi}}
\newcommand{\etavec}{\bm{\eta}}
\newcommand{\cvec}{\bm{c}}
\newcommand{\rvec}{\mathbf{r}}
\newcommand{\kvec}{\mathbf{k}}
\newcommand{\Kvec}{\mathbf{K}}
\newcommand{\Kperp}{\bm{K}_\perp}
\newcommand{\pc}{\bm{p}_{c}}

\begin{document}

\maketitle

\begin{abstract}
Associative recall—mapping an incident pattern to the stored one it most
resembles—is the natural computational primitive of a high-dimensional vision
front end, and it is precisely the operation a volume hologram performs
natively. We show that a cascade of two volume holograms separated by a
one-dimensional coded layer physically evaluates the modern Hopfield (dense
associative memory) retrieval map, $\hat{\eta}=V\,\mathrm{softmax}(\lambda
K^{\top}x)$, exactly as a parallel optical computation, with the inverse
temperature realized via optically addressed spatial light modulation in the coded-layer. Routing
the input and output through a 1D code rather than directly between 2D planes
supplies the separating nonlinearity the original Hopfield model lacked and, by
balancing the grating-wavevector dimension count ($2+1=3$), removes the Bragg
degeneracy that otherwise forces fractal sampling on a direct 2D-to-2D
hologram. Faithful dense storage further demands a recording medium that
captures inter-neuron connections while rejecting the field self-energy
responsible for the $M^{-2}$ efficiency falloff of homogeneous photorefractives.
We propose a nonlocal, gradient-responsive medium whose
illumination-independent decay recovers the linear $M^{-1}$ scaling in situ, and
demonstrate its reception, combination, and storage functions in a discrete
opposing-diode cell. Routes to OASLM-stack and volume molecular/nanocrystal
realizations are outlined.
\end{abstract}

\section{Introduction}
\label{sec:intro}

\begin{figure}
    \centering
    \includegraphics[width=0.75\linewidth]{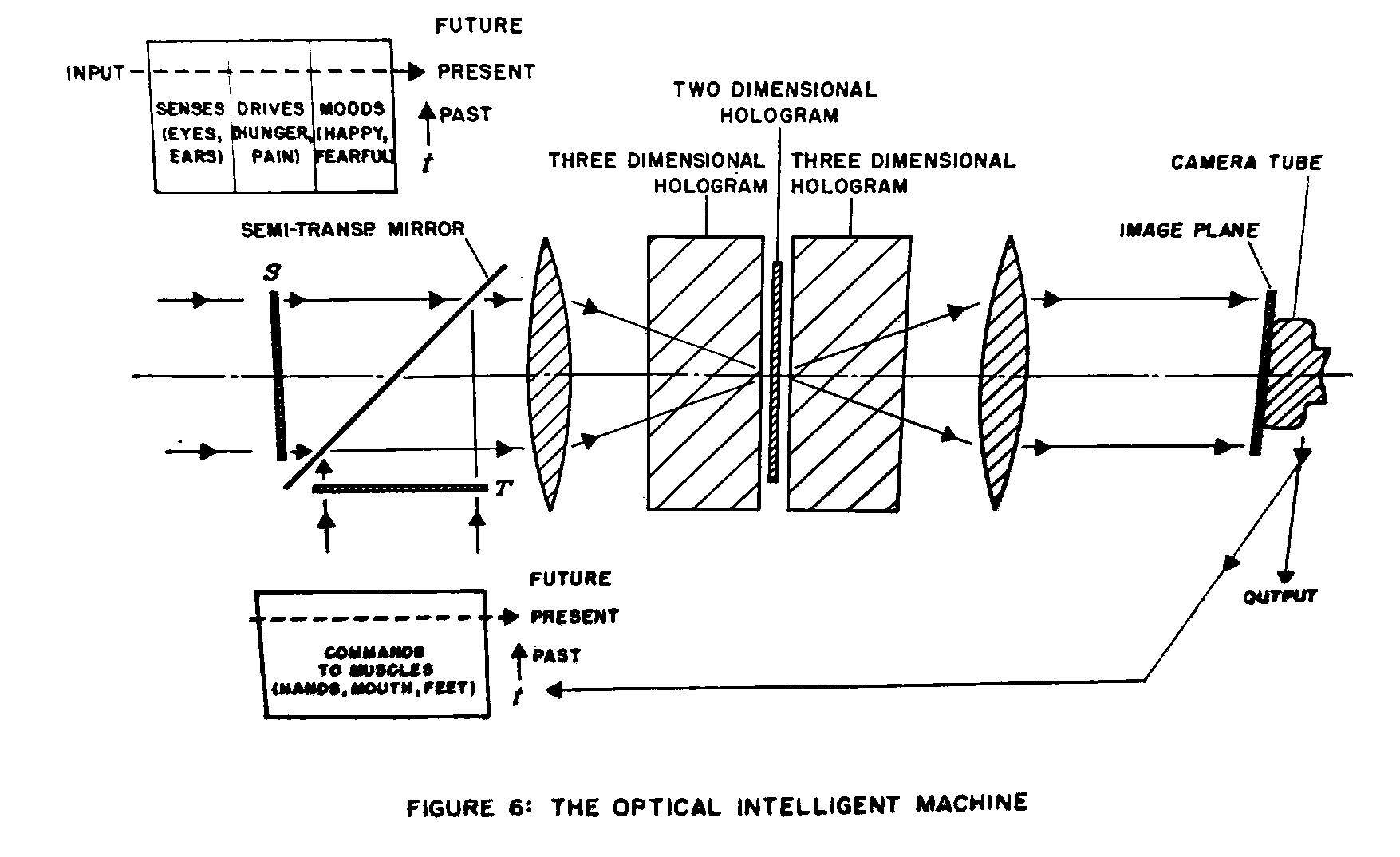}
    \caption{van Heerden's ``optical intelligent machine.'' Reprinted from~\cite{vanheerden1968foundation}.}
    \label{fig:vanHeerden}
\end{figure}

Van Heerden first proposed holographic associative memories over 60 years ago~\cite{van1963new, vanHeerden1963}. While van Heerden's proposal evolved over the course of a decade, it first appeared in corporate technical reports~\cite{vanheerden1961theory} that predated the development of off-axis~\cite{LeithUpatnieks1962} and volume~\cite{denisyuk1962reflection} holography. Van Heerden was inspired by Gabor~\cite{gabor1948new} and by Beurle's 1956 report suggesting that memory could be modeled by volume wave propagation~\cite{10.1098/rstb.1956.0012}. The original idea of holographic memory was based on the now familiar concepts that holographic interconnections model neural synapses and that holograms naturally recover full images from windowed inputs. This idea was subsequently explored in more detail by Gabor~\cite{5391791} and Longuet-Higgins {\em et al.}~\cite{longuet1970theories, willshaw1969non}. van Heerden's concept for a holographic processor is illustrated in Fig.~\ref{fig:vanHeerden}.

The reader will notice that Fig.~\ref{fig:vanHeerden} does not detail the function or composition of the various components shown. Here we revisit this architecture and discuss specifically how it can be used to implement modern Hopfield networks, which are also called dense associative memories~\cite{KrotovHopfield2016, Ramsauer2021, LucibelloMezard2024}. To achieve this objective, we propose novel volume holographic materials with a nonlocal recording response. Such materials enable the physical implementation of Hopfield networks without the recording artifacts that limit homogeneous materials~\cite{BradyPsaltis1992, psaltis1988adaptive, MokBurrPsaltis1996}. Prior to explaining these innovations, we briefly review the historical context and motivation for these devices. 

From a modern perspective, one is surprised that van Heerden~\cite{vanHeerden1970} and Willshaw {\em et al.}~\cite{willshaw1969non} could publish such speculative ideas in Nature. The device proposed in Fig.\ref{fig:vanHeerden} was not feasible when it was proposed because  solid state focal planes, spatial light modulators, dynamic holographic materials, and large scale integrated circuits had not yet been developed. Once these technologies became available, Psaltis {\em et al.}~\cite{PsaltisFarhat1985,farhat1985optical}, Anderson~\cite{Anderson1986}, Yariv {\em et al.}~\cite{Yariv1986} and Owechko {\em et al.}~\cite{Owechko1987} demonstrated holographic associative memories. Most of these systems relied on variations of Hopfield's associative memory model~\cite{Hopfield1982} and used photorefractive materials~\cite{Gunter1988} as recording media. 

Van Heerden dedicated {\em The Foundation of Empirical Knowledge}~\cite{vanheerden1968foundation} ``to the true scientists, who do not hold anything to be impossible, but do not believe it until they see it with their own eyes.''  This is a curious dedication for a theoretical text, but it is consistent with the many practical issues that arise in actual implementations of holographic associative memories. Online learning in such devices is particularly challenging.  Hopfield theory and gradient descent~\cite{Wagner1987} training methods provide some guidance, but these processes must be mapped onto the topology and dynamics of holographic materials. Holographic associative memories implement transformations between physical distributions; fractal sampling grids or other spatial filters must be applied to avoid unwanted crosstalk in such transformations~\cite{psaltis1988fractal}. More critically, dynamic holographic materials must balance the processes of recording, storage, and read-out~\cite{psaltis1988adaptive, BradyPsaltis1992}. Despite these challenges, several studies successfully demonstrated simple associative learning in photorefractive systems~\cite{Psaltis1990, BradyThesis1990, brady1988perceptron}. 

Holographic memory research largely shifted to random access designs in the 1990s~\cite{PsaltisBurr1998}. This shift coincided with a quiet period for the larger field of neuromorphic computing. As with the delay between proposals in the 1960s and experiments in the 1980s, the gap from 1990 to 2015 was due to electrical hardware deficiencies that were resolved by the development of graphical processing units~\cite{lecun2015deep}. With the recent success of deep AI models, interest in optical processors has surpassed previous levels~\cite{fu2024optical, wetzstein2020inference}. Recent work includes the use of optical nonlinearity in modern Hopfield systems~\cite{Li:26, Katidis:25}, but holographic associative memory research has not yet returned to previous intensity. In our opinion, this is because most current AI models are built around language processing, with nearly serial input signals. Demand for holographic associative memories will naturally arise in the development of high-dimensional vision processors. Just as current neural processors were not possible without significant hardware innovations, holographic associative memories will require major innovations in (1) optically addressed spatial light modulators (OASLM) and (2) active holographic materials. While OASLM are not currently commercially available with specifications consistent with high performance associative memories, there is substantial precedent for their construction. Active volume holographic materials, on the other hand, have no previous precedent. For this reason, design and analysis of these materials is a primary focus of this paper. 

Associative memory processes spatial or spatio-temporal patterns. Such processors are fundamentally related to visual media. Current visual media processing begins, unfortunately, with serial data streams. The basic function of the modern camera is to serialize two-dimensional optical fields into  one-dimensional
 digital streams. In doing so, it imposes three quantizations that
the field itself does not possess: a fixed pixel count $N^2$, a fixed frame rate
$f$, and a fixed bit depth $b$. The serial readout must carry the field through a
small number of physical lanes, and it is this channel—not the
photodetection—that sets the achievable temporal resolution and dominates the
sensor's power budget. This bottleneck is unfortunate; the image field is naturally spatially parallel. This field is initially processed by lens systems to form a coarse image. Lenses reasonably resolve up to $10^7$ spatial modes without aberration; parallel processing using multiscale lenses and integrated microcamera arrays enable cameras that exceed $10^9$ pixels per cm$^2$ of aperture~\cite{Brady2012, Brady2025Computational}. Transforming this massively parallel data stream into a serial signal requires discrete sampling and massive temporal downsampling. 
Frame rate is a construct of the readout. A frame exists only because
a serial channel must quantize time into the interval needed to scan the array
once; there is no frame in the optical field. One would like to register events
at the microsecond scale, but these occur at isolated locations, and a serializer can
capture them only by reading the entire array at the event timescale—an absurd
bandwidth demand for a signal that is almost everywhere static. Event-driven
sensors are often presented as an escape from the frame model, yet each pixel still
arbitrates onto a shared address--event bus, replacing periodic serialization
with asynchronous serialization without removing the single channel or its
bandwidth ceiling under high activity~\cite{Lichtsteiner2008}. Bit depth is the
same pathology at the level of amplitude: a uniform digitization pipeline assigns
every pixel the same number of bits, whereas the information available at a given
location is set by local scene statistics—nearero across a flat region, high
at an edge or a moving feature.

Biological vision does not operate under these constraints. The retina decorrelates the incident field
and channelizes it through on the order of $10^6$ parallel outputs, computing
contrast, edge, and motion in place and preserving spatial topology rather than
entropy-coding into a serial symbol stream. The visual front end is never reduced
to a single channel, and the quantities a camera fixes globally are, in the
retina, local and signal-dependent. This paper proposes an artificial front end
built on the same principles: optically addressed spatial light
modulators at the focal plane that transform the field where it lands, followed
by coherent volume-holographic processors that associate the transformed field
with stored patterns, with no intervening serial channel.

The three axes considered so far—space, time, and amplitude—substantially undercount
what serialization discards. The optical field carries further degrees of
freedom at every point: a spectral density, a polarization state, a depth or
focal structure, and a coherence structure, the mutual coherence function that a
square-law detector destroys by recording only time-averaged intensity~\cite{brady2025interferometric}. A
conventional sensor integrates over all of these to produce one real scalar per
pixel, so serialization does not merely quantize three axes—it projects an
entire optical data cube onto a single intensity image before quantizing what
survives. That the discarded axes are recoverable is not speculative: snapshot
compressive imaging codes a high-dimensional ($\ge\!3$D) cube into a single 2D
measurement and reconstructs it computationally, and it has been demonstrated across
the spectral, temporal, focal, polarization, and coherence
dimensions~\cite{YuanSCI2021}.

\begin{figure}
    \centering
    \includegraphics[width=0.8\linewidth]{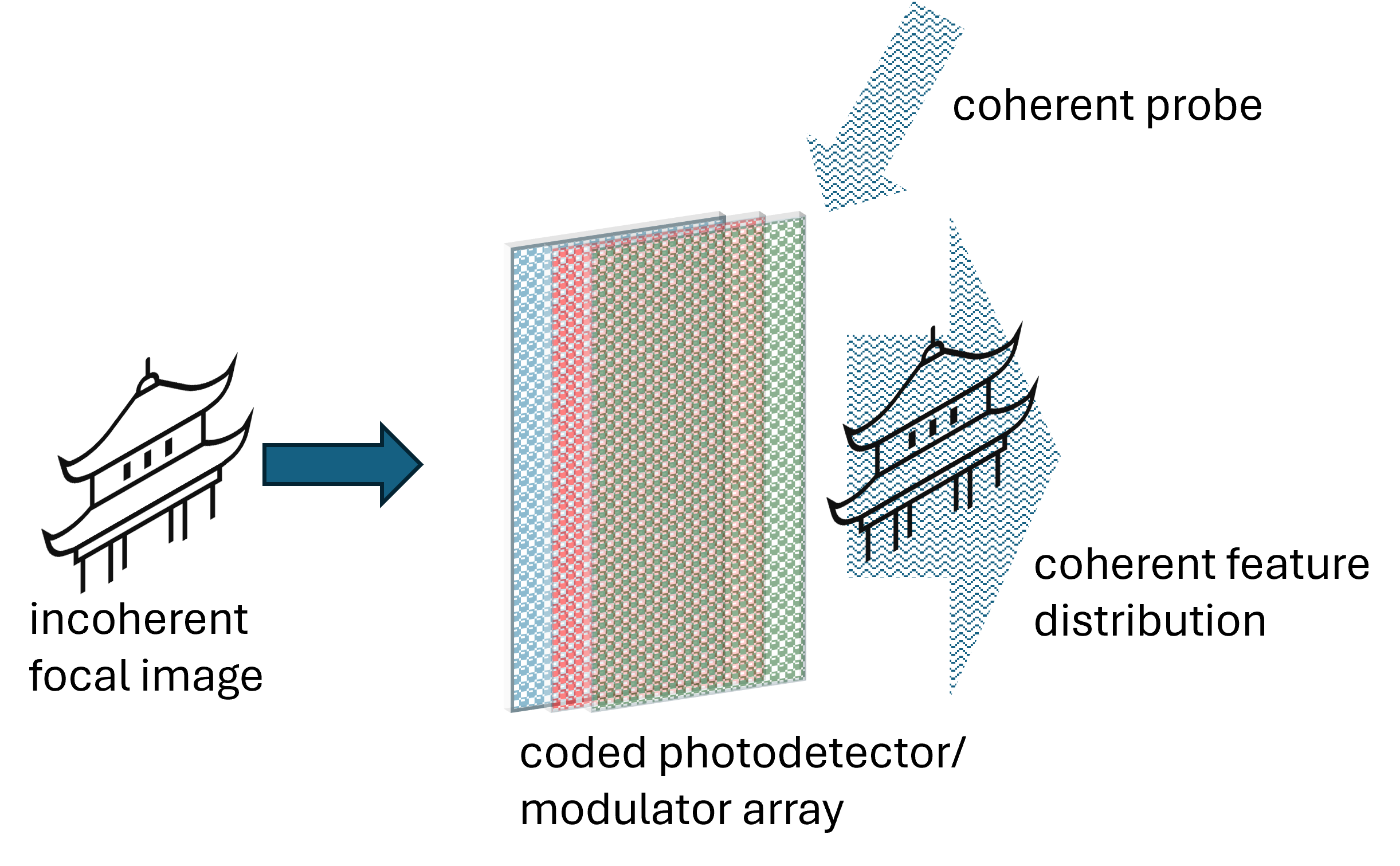}
    \caption{Optically addressed focal plane. The focal distribution generated by a natural light imaging system is incident from the left. The focal plane contains diverse sensors for irradiance, polarization, spectral and coherence features. Rather than electrical read-out, the transducer array modulates a coherent probe field. This modulated field propagates to the right as a 2D distribution. }
    \label{fig:opticalFocalPlane}
\end{figure}

The data cube-sampling problem and the associative-recall problem are properly
separated. Sensitivity to spectral, polarization, focal, and coherence content is
the job of \emph{the measurement interface}---a focal-plane stage of filters and
sampling structures (dichroics, polarizers and waveplates, microlens or
coded-aperture arrays, depth-coded masks, and the like) that imprints the
selected axes of the cube onto a two-dimensional pattern of optical amplitude and
phase. The natural realization of such an interface is an optically addressed
spatial light modulator at the focal plane: a transducer whose local optical
state is set by the incident field and read out, in parallel, by a coherent probe
beam. Devices of this kind have been developed since the
liquid-crystal-light-valve~\cite{Bleha1978} and, in the modern form most relevant
here, as hybrid amorphous-silicon photodiode/ferroelectric-liquid-crystal 
OASLMs achieving high resolution at video rates and beyond~\cite{Barbier1992,
Landreth1992}. Combined with data cube-sampling structures upstream, an OASLM acts as
the encoder: it converts the multi-axis natural field into
the unstructured two-dimensional coherent field that the rest of the system
reads. Such an OASLM is shown conceptually in Fig.~\ref{fig:opticalFocalPlane}. We are not  concerned with the details of such devices here, rather we focus on downstream processing of the coherent 2D signal. 

The processing layers downstream—the volume holograms and the coded layer
developed below-- operate on the 2D
coherent field handed off by the interface and have no spectral, polarization, or
focal sensitivity of their own. They use optical signals not because of any
intrinsic property of the coherent field, but because optics is the appropriate
platform for large-scale parallel data processing: it routes many channels in
parallel without a fan-out problem, supports up to $\sim\!10^{12}$ pairwise
connections without the wiring and dissipation that any
electronic substrate would impose at the same scale, and propagates without
$I^{2}R$ loss. The data cube axes a reviewer might be tempted to attribute to the
holographic substrate live, in this architecture, only at the interface.

The natural computational primitive for such a front end is associative
recall—mapping an incident pattern to a stored one. Modern Hopfield networks,
or dense associative memories (DAMs), have made this primitive precise: their
retrieval step is the attention update, a softmax-weighted combination of stored
values keyed by the correlation of a probe with stored
patterns~\cite{KrotovHopfield2016,Ramsauer2021,LucibelloMezard2024}.
Li {\em et al.}~\cite{Li:26} and Katidis {\em et al.}~\cite{Katidis:25} implement the nonlinear operation optically on electrically stored patterns. Here we assume that the softmax nonlinearity is implemented electro-optically and focus instead on holographically stored patterns. Associative pattern correlation is precisely the operation volume holography performs natively, since
up to $V/\lambda^3$ distinguishable plane-wave gratings can coexist in a medium
of volume $V$~\cite{vanHeerden1963,Psaltis1990}.

Previous holographic associative memories have not fulfilled the promise of
this approach for two related reasons: they were designed to implement the
\emph{original} Hopfield model rather than its modern, dense form, and even
that original model was not faithfully realized by the recording physics. We consider these issues in turn.
The original Hopfield network~\cite{Hopfield1982} writes the synaptic matrix
as the outer-product sum
\begin{equation}
    w_{ij}=\sum_{k} \xi^{(k)}_{i}\eta^{(k)}_{j}
\end{equation}
and retrieves by a single linear
projection $\hat{\eta}=Wx$, optionally followed by a pointwise threshold and
iteration. Storage capacity is only $\sim 0.14 N$ patterns in $N$ neurons, and
because retrieval is linear, non-orthogonal stored inputs interfere on
read-out~\cite{Psaltis1990}. Modern Hopfield networks~\cite{KrotovHopfield2016,Ramsauer2021}
replace the linear projection with a softmax-weighted attention update
—a separation-of-patterns nonlinearity that lifts capacity to exponential in $N$
and is exactly the operation performed by transformer
attention~\cite{Ramsauer2021}. The two models share the same memory matrices;
they differ only in the read-out nonlinearity, and that nonlinearity is what
turns overlap-induced crosstalk into clean retrieval. A single volume hologram
mapping the input plane directly to the output plane realizes only the
original, linear read-out: it has no place to apply the separating
nonlinearity, and non-orthogonal inputs blur on retrieval regardless of the
dimensionality of the input and output.

Holographic recording physics imposes a second, independent limitation. Learned
associations in photorefractive media are not purely proportional to the outer
product of the recording fields: the self-energy of each field also contributes
to the recorded hologram. The index modulation is supported by a finite,
\emph{shared} population of charge, so each new exposure redistributes this
population and partially erases its predecessors. The diffraction efficiency
of each of $M$ superimposed holograms therefore falls as
$M^{-2}$~\cite{Psaltis1990,MokBurrPsaltis1996}, rather than the $M^{-1}$ that
a fully utilized linear medium would allow. This is the limitation most
directly responsible for the factor-of-hundred gap between demonstrated
photorefractive memories~\cite{Mok1993} and the information-theoretic ceiling.

This paper proposes a holographic dense associative memory that is free of
both obstacles. Our first contribution (Sec.~\ref{sec:arch}) is to show that a \emph{multilayer}
volume hologram with a hidden layer between the input and output planes physically
evaluates the modern Hopfield/DAM retrieval map. The first hologram maps
each stored input $\xi^{(k)}$ to an orthogonal code $c^{(k)}$ on the hidden
layer; a pointwise nonlinearity on that layer implements the softmax, with the
amplifier gain serving as the inverse temperature $\lambda$; a second hologram
maps the activated code back to the stored output $\eta^{(k)}$. The hidden
layer is the architectural element that supplies the separating nonlinearity
that the original Hopfield formulation lacked—this is what promotes the device
from a linear correlator to a dense associative memory. We implement the
hidden layer as a one-dimensional coded layer, which allows the input and
output planes to retain the natural two-dimensional spatial structure of
images and, as a side benefit, eliminates the Bragg degeneracy that arises
when 2D-to-2D pairings are forced through the three available dimensions of
grating wavevectors~\cite{Lee1989}.
Our second contribution (Sec.~\ref{sec:medium}) is an in-situ distributed
recording medium whose cells respond to the spatial \emph{gradient} of the
recording intensity. This medium stores the inter-neuron connections while
rejecting the self-energy that produces the $1/M$ falloff. We argue that the
essential property of such a medium is nonlocality and identify the four
functions each cell must perform.

Biological visual processing is extraordinarily complex: the
retina alone performs substantial in-place computation before any signal
reaches the brain, and the cortical stages that follow are hierarchical,
massively recurrent, and still only partly understood. We do not model that
system, and we do not claim a complete architecture for machine vision. Our aim
is deliberately narrower. Recognizing the renewed interest in optical
processors for large-scale vision, we show that a holographic substrate can
evaluate the modern Hopfield (dense associative memory) retrieval map exactly,
as a physical computation, and can do so at a density of parallel weighted
interconnections—of order $10^{12}$ connections in a centimeter-scale
volume. Associative recall is, in this sense, the operation for which
volume holography is the natural hardware. The remainder of the paper makes
this concrete: section~\ref{sec:arch} develops the two-stage optical
architecture, section~\ref{sec:medium} describes the recording dynamics that overcome M-number limitations, and
section~\ref{sec:cells} describes device and materials design to achieve the correct spatial and temporal recording response. 
While the analysis presented here makes van Heerden's original concept considerably more concrete, we recognize that a dense holographic associative memory 
will be just one component in visual processing systems. Section~\ref{sec:roadmap} briefly considers the roadmap
for continuing development. 

\section{Dense associative memories using multilayer holograms}
\label{sec:arch}

In this section, we show that cascaded volume holograms separated by a one-dimensional coded layer, with a nonlinearity
applied at that layer, physically evaluate the retrieval map of a dense associative memory. We also explore the physical layout and network capacity of this architecture. Multilayer volume holographic associative networks were originally proposed by Wagner and Psaltis~\cite{Wagner1987}, but their report focused on gradient descent learning. Here we consider multilayer networks with Hopfield learning.
A modern Hopfield network stores $M$ associations
$\{(\xivec^{(k)},\etavec^{(k)})\}_{k=1}^{M}$, with inputs $\xivec^{(k)}\in
\mathbb{R}^{N}$ and outputs $\etavec^{(k)}\in\mathbb{R}^{N}$. Given a probe
$\bm{x}$, recall is one step of the attention
map~\cite{Ramsauer2021,LucibelloMezard2024}
\begin{equation}
\hat{\etavec} = \sum_{k=1}^{M} a_k\,\etavec^{(k)},
\qquad
a_k = \frac{e^{\lambda\,\langle \xivec^{(k)},\bm{x}\rangle}}
{\sum_{j=1}^{M} e^{\lambda\,\langle \xivec^{(j)},\bm{x}\rangle}},
\label{eq:attention}
\end{equation}
i.e. $\hat{\etavec} = V\,\mathrm{softmax}(\lambda\,K^{\!\top}\bm{x})$ with key
matrix $K=[\xivec^{(1)},\dots,\xivec^{(M)}]$ and value matrix
$V=[\etavec^{(1)},\dots,\etavec^{(M)}]$. The scalar $\lambda$ is the inverse
temperature. As $\lambda\to\infty$ the softmax collapses to an $\arg\max$ and
\eqref{eq:attention} returns the single stored output whose input is most
correlated with the probe; as $\lambda\to0$ the weights flatten, and the output is
the barycentre of all stored values. The intermediate regime, where the weights
$a_k$ are graded, is what distinguishes a dense associative memory from a look-up
table: it is responsible for the network's basins of attraction and for the soft
blending of stored items~\cite{LucibelloMezard2024}. The auto-associative case
$\etavec^{(k)}=\xivec^{(k)}$ recovers the usual content-addressable memory;
keeping $\etavec^{(k)}\neq\xivec^{(k)}$ gives the more general hetero-associative
(cross-attention) map, which the optical device realizes with no additional cost.

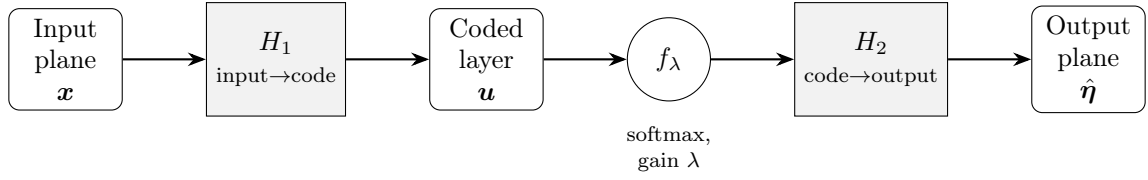
\begin{figure}[t]
\centering
\begin{tikzpicture}[
  node distance=7mm and 11mm,
  box/.style={draw,rounded corners,minimum height=11mm,minimum width=15mm,
              align=center,font=\small},
  holo/.style={draw,fill=black!5,minimum height=15mm,minimum width=14mm,
               align=center,font=\small},
  nl/.style={draw,circle,minimum size=11mm,align=center,font=\small},
  arr/.style={-{Stealth[length=2.4mm]},thick}
]
\node[box] (X) {Input\\plane\\$\bm{x}$};
\node[holo,right=of X] (H1) {$H_1$\\\scriptsize input$\to$code};
\node[box,right=of H1] (C) {Coded\\layer\\$\bm{u}$};
\node[nl,right=of C] (f) {$f_\lambda$};
\node[holo,right=of f] (H2) {$H_2$\\\scriptsize code$\to$output};
\node[box,right=of H2] (Y) {Output\\plane\\$\hat{\etavec}$};
\draw[arr] (X) -- (H1);
\draw[arr] (H1) -- (C);
\draw[arr] (C) -- (f);
\draw[arr] (f) -- (H2);
\draw[arr] (H2) -- (Y);
\node[below=2mm of f,font=\scriptsize,align=center] {softmax,\\gain $\lambda$};
\end{tikzpicture}
\caption{Architecture of the two-stage holographic dense associative memory. The
first hologram $H_1$ associates the 2D input with a code along the 1D coded layer;
the coded-layer softmax of gain $\lambda$ (the inverse temperature) restores a
clean code; the second hologram $H_2$ reconstructs the 2D output. The cascade
$H_2\circ f_\lambda\circ H_1$ equals $V\,\mathrm{softmax}(\lambda K^{\!\top}x)$.}
\label{fig:arch}
\end{figure}

As illustrated in Fig.~\ref{fig:arch}, the proposed architecture comprises three neural planes: a two-dimensional
input plane carrying the probe; a one-dimensional coded layer; and
a two-dimensional output plane. Two volume holograms connect the neural layers. The first,
$H_1$, is recorded so that each association couples its input $\xivec^{(k)}$ to a
code pattern $\cvec^{(k)}$ along the coded layer; the second, $H_2$, couples the
same code $\cvec^{(k)}$ to the output $\etavec^{(k)}$. With the codes written as a
distributed amplitude-and-phase modulation along the layer, the recorded
interconnects are
\begin{equation}
W^{(1)}_{ai} = \sum_k c^{(k)}_a \xi^{(k)}_i, \qquad
W^{(2)}_{ja} = \sum_k \eta^{(k)}_j c^{(k)}_a,
\label{eq:W}
\end{equation}
where $i$ indexes input pixels, $a$ the coded-layer channels, and $j$ output
pixels. Reading $H_1$ with a probe $\bm{x}$ produces the coded-layer field
$u_a=\sum_i W^{(1)}_{ai} x_i = \sum_k c^{(k)}_a \langle\xivec^{(k)},\bm{x}\rangle$.
The coded layer is a one-dimensional photonic processor; it
recovers the coefficient of each code by correlation against the
(well-conditioned) code bank,
\begin{equation}
\alpha_k = \langle \cvec^{(k)},\bm{u}\rangle = \langle\xivec^{(k)},\bm{x}\rangle,
\label{eq:correlate}
\end{equation}
the last equality holds because the codes are orthonormal. It then applies the
softmax of gain $\lambda$ to the coefficients and re-emits the restored coded
field $\bm{u}' = \sum_k a_k \cvec^{(k)}$, with $a_k$ as in \eqref{eq:attention}.
Reading $H_2$ with $\bm{u}'$ gives
\begin{equation}
\hat{\eta}_j = \sum_a W^{(2)}_{ja} u'_a
= \sum_a \Big(\sum_k \eta^{(k)}_j c^{(k)}_a\Big)
            \Big(\sum_{k'} a_{k'} c^{(k')}_a\Big)
= \sum_k a_k \eta^{(k)}_j,
\label{eq:cascade}
\end{equation}
using $\sum_a c^{(k)}_a c^{(k')}_a = \delta_{kk'}$. Equation~\eqref{eq:cascade} is
identical to the attention update \eqref{eq:attention}: the cascade
$H_2\circ f_\lambda\circ H_1$ is the DAM retrieval map, with the coded-layer
softmax $f_\lambda$ supplying the only nonlinearity.

The hidden layer nonlinearity would most easily be realized in a 1D optically addressed spatial light modulator. The first holographic layer correlates the 
input distribution over the stored patterns, and then the hidden layer detects the coded-layer field $u_a$. This field signal is electronically processed and 
then drives coherent modulators to produce restored coded field $u'$ that drives the second holographic layer to retrieve the memory. While one may imagine implementing this operation optically, as described by Li {\em et al.}~\cite{Li:26} and Katidis {\em et al.}~\cite{Katidis:25}, electro-optical processing is well-established and much simpler to implement. In any case, we do not consider the details of the nonlinear layer here; rather, we focus on the system geometry and recording mechanism. 

The code layer is specified as a 1D signal to address well known Bragg degeneracies that would otherwise create model error. 
In the simplest case, each interconnect is a grating between input and output plane waves. 
This is the limit in which holographic interconnection was originally conceived, and the limit to which the well-known
degeneracy analysis of Lee, Gu and Psaltis~\cite{Lee1989} applies.
A plane-wave component with wavevector $\kvec$ is constrained to the wavelength sphere
$|\kvec| = 2\pi/\lambda$ and so carries two angular degrees of freedom. A 2D SLM in
the front focal plane of a Fourier lens excites a 2D subset of this sphere (2 DOF);
a 1D arrangement of pixels along one axis excites a 1D subset (1 DOF). Recording by
interfering an input field with an output (or code) field generates index gratings
whose wavevectors are differences $\Kvec = \kvec_{\mathrm{out}} -
\kvec_{\mathrm{in}}$. Unlike the optical wavevectors, $\Kvec$ is not constrained to
a sphere: it is an arbitrary 3D vector limited only by the $2k_0$-ball of the Ewald
construction, and so carries up to 3 degrees of freedom. At readout, a probe
$\kvec_{\mathrm{probe}}$ excites a diffracted wave $\kvec_{\mathrm{out}} =
\kvec_{\mathrm{probe}} + \Kvec$ only when $|\kvec_{\mathrm{out}}| =
|\kvec_{\mathrm{probe}}|$; this Bragg condition selects which gratings contribute
for a given probe.

Let $d_{\mathrm{in}}$ and
$d_{\mathrm{out}}$ be the dimensionalities of the input and output wavevector
distributions, and $d_{\Kvec}=3$. The map
$(\kvec_{\mathrm{in}},\kvec_{\mathrm{out}})\mapsto\Kvec$ takes a
$(d_{\mathrm{in}}+d_{\mathrm{out}})$-dimensional source to a 3-dimensional target.
When $d_{\mathrm{in}}+d_{\mathrm{out}}>3$ the map is overdetermined: a continuous
family of pairs shares each $\Kvec$, and the hologram cannot distinguish them. A
single 2D-to-2D layer has $2+2=4>3$, leaving one degree of forced degeneracy; this
is the geometric origin of single-layer crosstalk, removable only by fractal
sampling that breaks the natural structure of image data~\cite{psaltis1988fractal}. That a direct 2D-to-2D map is unusable is not in dispute; the point here is that the
code-mediated architecture removes the difficulty for free.

\begin{figure}[t]
\centering
\includegraphics[width=0.62\linewidth]{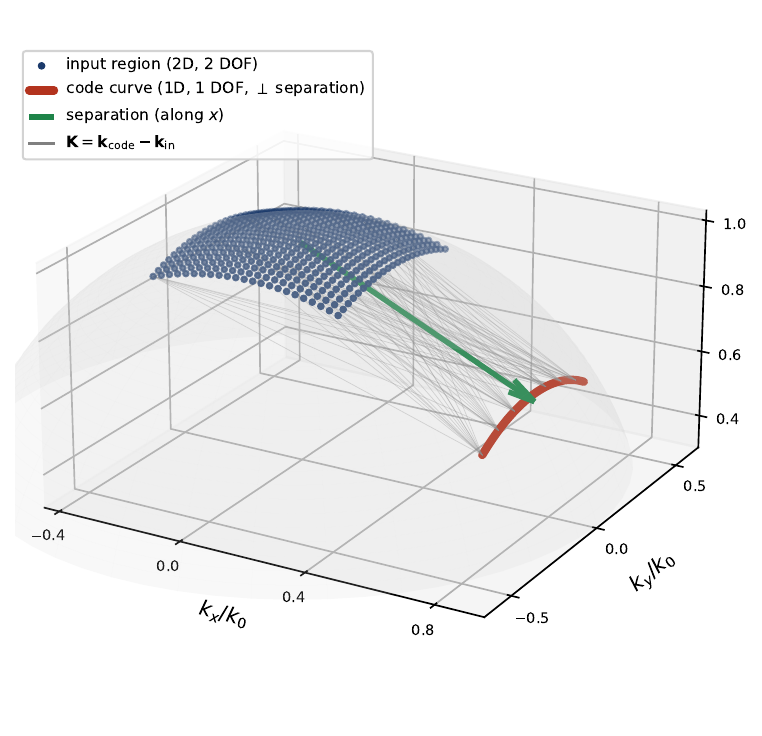}
\caption{First-layer geometry on the wavelength sphere. The two-dimensional input
region (2~DOF) and the one-dimensional code curve (1~DOF), oriented orthogonal to
their separation so the beams do not overlap on the sphere; grey lines are
representative connecting grating vectors
$\Kvec=\kvec_{\mathrm{code}}-\kvec_{\mathrm{in}}$. The pair count
$2+1=3=d_{\Kvec}$ makes the
$(\kvec_{\mathrm{in}},\kvec_{\mathrm{code}})\mapsto\Kvec$ map a bijection onto its
image, removing the single-layer Bragg degeneracy.}
\label{fig:ewald}
\end{figure}

The code-mediated architecture uses two dimension-mismatched layers
(Fig.~\ref{fig:ewald}). The first maps the 2D input to the 1D code axis,
$2+1=3=d_{\Kvec}$; the second maps the 1D code axis back to the 2D output,
$1+2=3$. In both cases, the map
$(\kvec_{\mathrm{in}},\kvec_{\mathrm{code}})\mapsto\Kvec$ is generically a
bijection onto its image, and the degeneracy disappears. Geometrically, the input
populates a two-dimensional cap of the wavelength sphere, while the code occupies a
one-dimensional curve oriented orthogonally to the input--code separation, so the
two beam bundles do not overlap; each stored connection is the difference vector
$\Kvec=\kvec_{\mathrm{code}}-\kvec_{\mathrm{in}}$ joining a point of the code curve
to a point of the input cap. As $\kvec_{\mathrm{in}}$ sweeps the cap and
$\kvec_{\mathrm{code}}$ the curve, these difference vectors fill a
three-dimensional set, so distinct input--code pairs are carried to distinct
gratings. Two consequences follow: continuous image fields map without the
artificial sampling that fractal grids require, and the one-dimensionality of the
code is not merely convenient but geometrically required—it is what balances the
dimension count. In this idealized global limit, the available grating wavevectors
carry independent information, so capacity is structurally maximal; the practical
capacity of a buildable device is set by volume Bragg selectivity.

Plane-wave interconnections require a Fourier transform between neural layers. This requirement drew an early
and influential objection. van Heerden's proposal that memory is a
volume-holographic transformation implies that each synapse performs a global
operation on the whole input field; Willshaw, Buneman, and Longuet-Higgins objected
that biological connectivity is sparse and local and exhibited a simple
non-holographic associative net that performs the same recall with local
connections alone~\cite{willshaw1969non}. van Heerden replied that the two views are
not in conflict~\cite{vanHeerden1970}, but the practical question
—whether holographic recall \emph{requires} global synapses—remained open.

This issue is resolved by assuming that the holographic transformations rely on \emph{shift
multiplexing}~\cite{Psaltis1995shift,Barbastathis1996}: the reference is a
spherical wave, and successive associations are distinguished not by reference
angle but by a relative translation of the medium. The one-dimensional coded layer
is then physically a shift axis—a line of spherical-reference foci—along which
the $M$ associations are written at distinct shifts, each separated from its
neighbors by the volume shift selectivity. Because the connection a
shift-multiplexed hologram implements is shift-invariant with local support, the
device moves away from the global Fourier transform and toward the locally-coupled,
propagating picture of memory that Beurle proposed~\cite{10.1098/rstb.1956.0012}: recall is
carried by local interactions rather than by an all-to-all transform. Locality at
the interconnect and clean separation of associations are therefore not in tension.
The orthogonality that separates associations lives on the code axis,
while the connections that feed it are as local as the recording geometry makes
them— which is the concrete sense in which the device meets the
Willshaw objection without abandoning the holographic substrate.

\begin{figure}[t]
\centering
\includegraphics[width=\textwidth]{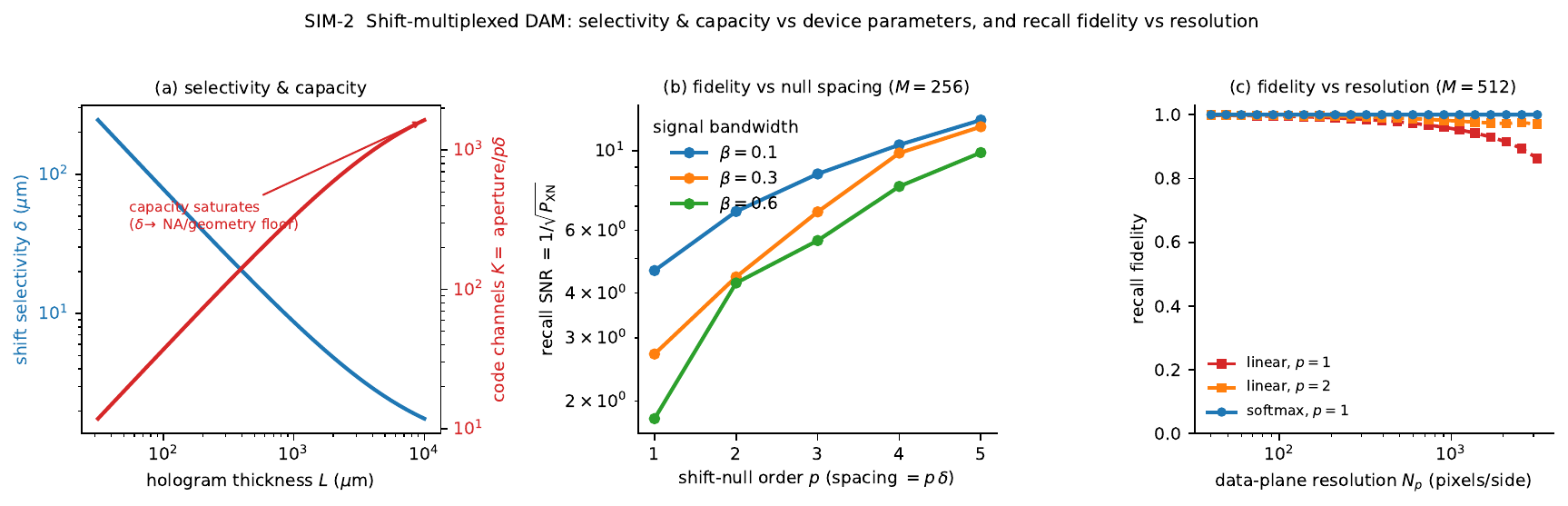}
\caption{Shift-multiplexed realisation: design parameters, capacity, and recall
fidelity, evaluated with the selectivity and crosstalk theory
of~\cite{Barbastathis1996}. (a)~Shift selectivity $\delta$ (left axis) falls with
thickness toward an NA/geometry floor, so the number of code channels
$C=s/(p\delta)$ (right axis) rises and saturates---the thickness--capacity
trade-off of a real volume. (b)~Recall SNR versus shift-null order $p$ for three
signal bandwidths $\beta$ (a resolution proxy): spacing at higher nulls and lower
bandwidth both improve fidelity, and the curves are essentially independent of the
stored load $M$. (c)~Recall fidelity versus data-plane resolution at fixed load: a
linear readout degrades as pages grow, while the code-layer softmax restores
near-ideal recall and so permits the densest ($p=1$) packing.}
\label{fig:sim2}
\end{figure}

Shift multiplexing fixes the device's capacity and fidelity through a small set of
physical parameters, which we evaluate using the selectivity and
crosstalk theory of Barbastathis, Levene, and Psaltis~\cite{Barbastathis1996}. The
shift selectivity—the translation between resolvable associations—is
\begin{equation}
\delta \;=\; \frac{\lambda_0\,z_0}{n_0 L\,\tan\theta_S'}
\;+\; \frac{\lambda_0}{2\,\mathrm{NA}},
\label{eq:shiftsel}
\end{equation}
with $z_0$ the apparent focal distance of the spherical reference, $\theta_S'$ the
internal signal angle, $n_0$ the index, and $L$ the thickness. The first term is
the volume Bragg term; the second is the finite-aperture spot ambiguity. Because
the recording geometry forces $z_0$ to grow with $L$, $\delta$ does not fall as
$1/L$ indefinitely but tends toward a floor, so the number of code channels at one
location, $C = s/(p\,\delta)$ for a signal aperture $s$ and null-order spacing $p$,
rises with thickness and then saturates (Fig.~\ref{fig:sim2}a). For representative
parameters ($\lambda_0=0.5\,\mu$m, $\mathrm{NA}=0.5$, $\theta_S=30^\circ$, second
null $p=2$), the code count $C$ is $40$, $300$, and $1600$ channels at $L=0.1$,
$1$, and $10$~mm. The data-plane resolution $N^2$ is an independent budget set by
the signal optics, so the information stored at one location scales as $N^2 C$. 
For a typical aperture-limited space-bandwidth product, $N^2$ will be 10-100 megapixels.

Recall fidelity is governed by interpage crosstalk. Associations spaced at the
$p$-th Bragg null are nulled only at the carrier; pixels away from the carrier
—equivalently, higher signal bandwidth—detune from the null and leak. Two design
knobs follow (Fig.~\ref{fig:sim2}b): spacing at higher nulls (larger $p$) suppresses
crosstalk at a proportional cost in $C$, and coarser pages (lower signal bandwidth)
are cleaner. The crosstalk is dominated by near shift-neighbors and saturates with
the number stored, so recall fidelity is essentially independent of the total load
$M$; unlike same-spot methods such as angle multiplexing, the stored load does not
itself degrade recall. Finally, the code-layer softmax sharpens the recovered
coefficients before $H_2$ reconstructs the output, and this is what licenses
aggressive packing: at the densest spacing ($p=1$) a purely linear readout degrades
as the data-plane resolution grows, whereas the softmax-restored recall remains
near-ideal (Fig.~\ref{fig:sim2}c). The nonlinearity that makes the device a dense
associative memory is thus also the nonlinearity that recovers the fidelity lost
to dense shift packing, so the capacity of \eqref{eq:shiftsel} can be used in full.

It is important to state precisely what the cascade does and does not establish.
We realize the DAM retrieval map \eqref{eq:attention} — the attention update, with
a hardware-tunable temperature — exactly as a physical computation. We do not
claim the exponential-in-$N$ capacity theorem of the random-pattern ensemble
analysis~\cite{LucibelloMezard2024}; that is a statement about how many random
patterns an energy function admits as attractors, and it holds for any device that
evaluates \eqref{eq:attention} at the cost of $O(M)$ stored patterns. The capacity
of the present device is set, as in any associative memory, by the number of
physically realizable modes: at most one association per orthogonal code, $M\le K$,
with the code count $C=s/(p\delta)$ fixed by the shift selectivity
\eqref{eq:shiftsel} and ultimately bounded by the $N^2 C \lesssim
(L/\lambda_{\mathrm{opt}})^3$ grating count of a hologram of edge $L$. This is the
same linear-in-$M$ resource scaling as the attention sum itself. The contribution
is therefore not a new capacity regime but (i) the physical realization of the
attention/DAM retrieval map in a parallel optical substrate, with the inverse
temperature as an amplifier gain, and (ii) the recording medium of
Sec.~\ref{sec:medium} that makes the underlying gratings dense by rejecting
self-energy in situ. Because the coded layer performs the softmax globally over
all code channels, the recall dynamics are independent of the choice of code; the
codes are free to be selected on recording grounds (low peak-to-average index
modulation; see Sec.~\ref{sec:medium}), the design freedom we exploit there.

\section{Recording dynamics}
\label{sec:medium}

Section~\ref{sec:arch} showed that a cascade of volume holograms separated by a
coded layer, $H_2\circ f_\lambda\circ H_1$, physically evaluates the modern
Hopfield (dense associative memory) retrieval map\emph{—provided} the two
interconnects store exactly the associations of Eq.~\eqref{eq:W}. Those
associations are cross-correlations between distinct neurons: the weight
$W_{ai}$ couples input mode $i$ to code mode $a$ and carries no contribution
from either mode in isolation. The recording medium must lay this down
faithfully, and naive holography does not. A medium that responds to the local
recording intensity stores, alongside each connection, the self-energy of the
interfering fields— a quantity that holds no relational information and only
consumes dynamic range. Superimposing $M$ associations then buries the
connections under an accumulating self-energy pedestal, and the stored matrix
departs from Eq.~\eqref{eq:W}. Recording the connections while rejecting the
self-energy is therefore a prerequisite for a dense memory, not an optimization.

This section describes a recording medium that satisfies this criterion. We show that the operation
isolating the connection is a spatial high-pass (gradient) response, and that a
gradient is intrinsically nonlocal, so that \emph{nonlocality} is the defining
structural property of the medium~(Sec.~\ref{ssec:gradient}). We then observe that
photorefractive recording \emph{already} supplies this gradient response; its
limitation lies not in the spatial response but in a decay whose rate is locked
to the recording intensity, so that uniform illumination erases the stored
grating. Replacing that decay with an autonomous, illumination-independent
forgetting yields a medium that records according to
Eq.~\eqref{eq:W} and recovers the linear $M^{-1}$
per-grating efficiency that a fully utilized medium allows.

\label{ssec:gradient}
When an input beam carrying $E_i e^{i\kvec_i\cdot\rvec}$ and a code beam carrying
$E_a e^{i\kvec_a\cdot\rvec}$ illuminate the medium together, the intensity is
\begin{equation}
I = \big|E_i e^{i\kvec_i\cdot\rvec} + E_a e^{i\kvec_a\cdot\rvec}\big|^2
  = \underbrace{|E_i|^2 + |E_a|^2}_{\text{self-energy (DC)}}
  + \underbrace{2\,\mathrm{Re}\big(E_i E_a^{*}\,
       e^{i(\kvec_i-\kvec_a)\cdot\rvec}\big)}_{\text{connection (grating)}}.
\label{eq:crossterm}
\end{equation}
The connection one wants to store---the weight $W_{ai}$---is the
cross-term: a fringe at wavevector $\Kvec_{ai}=\kvec_i-\kvec_a$. The self terms
are slowly varying offsets. A medium sensing the local intensity records both;
with $M$ associations superimposed, the self terms accumulate as a DC pedestal
growing with $M$, while each connection grows only as $\sqrt{M}$. Thus, each 
connection's share of the index range falls and the recorded matrix drifts away
from Eq.~\eqref{eq:W}~\cite{Dinc2024}.
The operation that separates the two is a spatial derivative. The self terms are
the low-frequency envelope; the connection is the high-frequency fringe. Sensing
$\nabla I$ rather than $I$ is a high-pass filter: it annihilates the DC pedestal
and passes the fringe, projecting onto the relational subspace and discarding the
self-subspace. 

The spatial derivative is a {\em nonlocal} operation. This is achieved in a physical material with a response generated over a distribution of points. In contrast to homogeneous photorefractive materials, we propose here inhomogeneous holographic materials consisting of 3D arrays of recording cells.
The minimal nonlocal cell is a single differential pair: two photo receptors separated
by a baseline $\pc$ whose response difference approximates the directional
derivative $\pc\cdot\nabla I$. This vanishes for uniform illumination and is
maximal for a fringe aligned with $\pc$; for a transverse grating $\Kperp$ its
response is $\propto\sin(\tfrac12\Kperp\cdot\pc)$, a high-pass kernel with an
exact null at $\Kperp=\bm 0$. The cell stores connections and ignores
self-energy by design. The purpose of the cell is to record signals arising from interneuron interference and to use the recorded signals to interconnect neurons. While one could imagine isolating the interneuron signals by other means, such as spectral, polarization, or temporal coding, the spatial gradient is a particularly simple and effective strategy for achieving this goal.

Media with exactly this gradient response already exists. In a photorefractive
crystal, the index grating is written by the space-charge field that mobile
photo-carriers build up, and in the diffusion-dominated, low-modulation regime
, that field is driven by the gradient of the recording
intensity~\cite{BradyPsaltis1992}:
\begin{equation}
\frac{\partial n(\rvec,t)}{\partial t}
= a\,\nabla I(\rvec,t) - b\,I(\rvec,t)\,n(\rvec,t).
\label{eq:pr}
\end{equation}
The recording term is already $a\,\nabla I$: photorefractive diffusion is a
high-pass response that vanishes at $\Kvec=\bm 0$, isolating the cross-term and
rejecting the self-energy exactly as the differential pair does. The cross-term
selectivity we require is thus not exotic—it is intrinsic to photorefractive
transport. The difficulty lies entirely in the second term. The same mobile
charge that gives the gradient response also relaxes the stored field at a rate
$bI$ set by the recording intensity, with two consequences. First,
\emph{read-erases-write}: a uniform read beam has $\nabla I=0$ and writes nothing,
but it still drives the decay, so a stored grating relaxes at
$\propto bI_{\mathrm{read}}$ while it is probed—constant illumination erases the
space-charge grating. No single-wavelength choice of parameters can make the
material plastic for writing and quiescent for reading, because one constant $b$
governs both. Second, each new exposure erases its predecessors at a rate set by
its own intensity, and the optimal equalized schedule for $M$ superimposed
gratings then yields per-grating efficiency $\propto M^{-2}$.
\ Both consequences, together with the depletion of one finite, shared charge
reservoir by successive exposures, trace to a single fact: storage is carried by
mobile charge, whose decay is locked to the illumination.

A medium that records according to Eq.~\eqref{eq:W} must keep the gradient
recording term and eliminate this illumination-locked decay. We propose a state
law
\begin{equation}
\;\frac{\partial n(\rvec,t)}{\partial t}
= a\,\nabla I(\rvec,t) - \frac{n(\rvec,t)}{\tau}\;
\label{eq:active}
\end{equation}
in which the relaxation time $\tau$ is a fixed material property of the cell's
internal storage channel, \emph{independent of} the recording intensity. Each
stored grating relaxes toward zero on its own autonomous clock, regardless of
what is being written or read elsewhere in the volume: the weights are neither
static nor erased by their own operation, but are continuously evolving on a chosen
timescale. This decouples writing from reading---a read beam, whether uniform or not, no
longer erases storage faster than the intrinsic $\tau$---and, because each cell's
state is its own and draws on no shared reservoir, storing one association no
longer depletes another. The two mechanisms by which photorefractives fall to
$M^{-2}$, the illumination-locked decay and the shared-reservoir depletion, are
removed together since both stem from mobile-charge storage, and the differential
cell replaces it with an autonomous per-cell state. The medium still forgets, but
by design rather than through interference in the act of being read.

The two terms of Eq.~\eqref{eq:active} can be controlled independently. A
dimensionless write gate $g(t)\in[0,1]$ in series with the recording term,
\begin{equation}
\frac{\partial n(\rvec,t)}{\partial t}
= g(t)\,a\,\nabla I(\rvec,t) - \frac{n(\rvec,t)}{\tau},
\label{eq:gated}
\end{equation}
licenses plasticity only when $g>0$ and leaves the forgetting rate untouched:
$g=1$ writes at full sensitivity, $g=0$ is pure recall. This is the device form
of the read-cycle/write-cycle separation that neuromodulatory gating provides in
cortex~\cite{Hasselmo2006}, and it is available precisely because the decay is not
illumination--locked—in a photorefractive medium, no such gate exists, since
reading necessarily erases. Because $\tau$ is set per cell, a medium may also
carry a spectrum of relaxation times at once, with fast-forgetting cells tracking the
most recent input and slow-forgetting cells holding a longer history.

\begin{figure}[t]
\centering
\includegraphics[width=\linewidth]{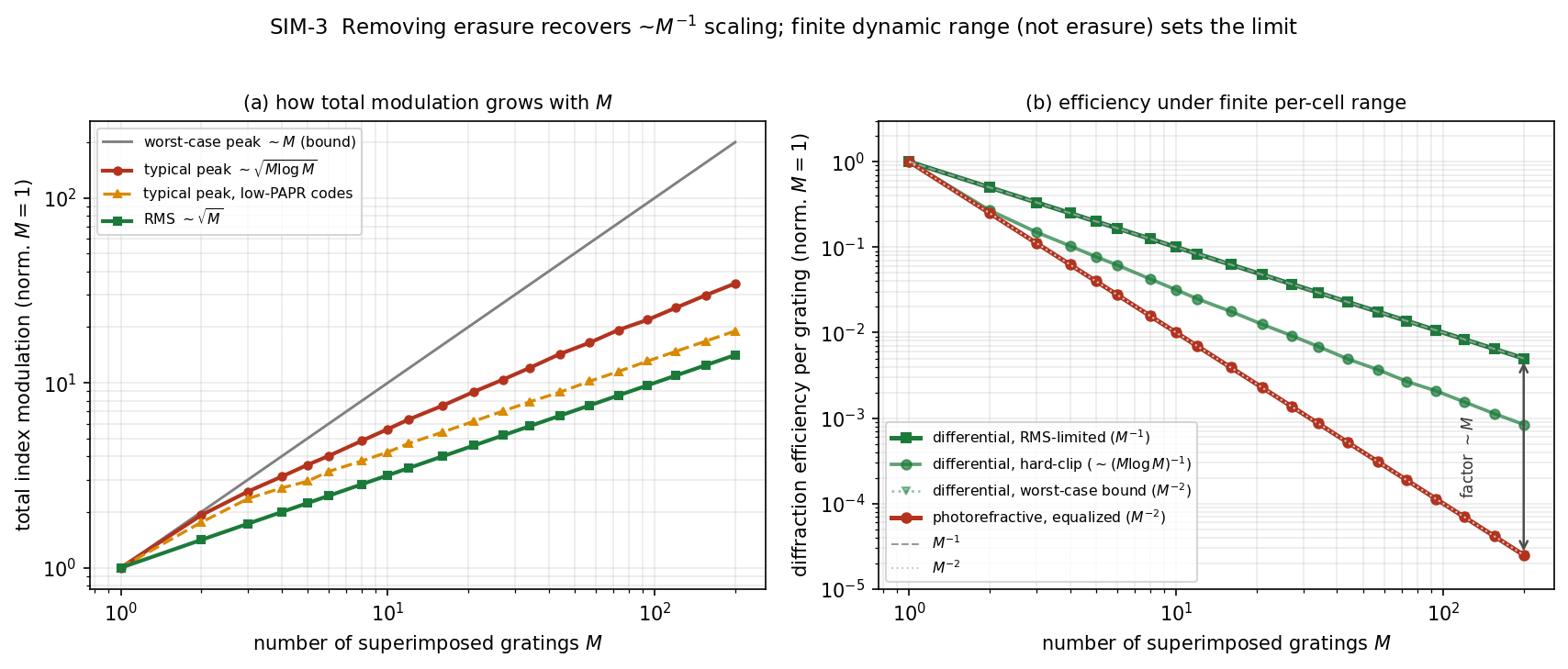}
\caption{Recovering the linear relation. (a)~The total recorded index modulation
grows as $\sqrt{M}$ (root-mean-square) to $\sqrt{M\log M}$ (typical peak), never
the worst-case $M$. (b)~Per-grating diffraction efficiency: the differential cell
tracks $M^{-1}$ (ideal linear medium) against the photorefractive $M^{-2}$, a
separation growing as $M$; low-peak code phases preserve $M^{-1}$ even under hard
clipping.}
\label{fig:sim3}
\end{figure}

With both erasure mechanisms removed, the limit on superimposed efficiency is set
by dynamic range alone. With the DC nulled, the index modulation in a cell is a
superposition of $M$ distinct-frequency gratings whose excursion has
a root-mean-square $\sqrt{M}$ and a typical peak $\sqrt{M\log M}$, the expected
maximum of the random superposition over the aperture; only the never-realized
case of all gratings momentarily in phase grows as $M$ (Fig.~\ref{fig:sim3}a).
Under a finite per-cell ceiling, the admissible per-grating amplitude, and hence
the efficiency, depends on which excursion saturates the cell: an average-power
limit gives the ideal $\eta\propto M^{-1}$; a hard clip at the typical peak gives
$\eta\propto(M\log M)^{-1}$, worse only by a logarithmic factor; only conservative
sizing against the worst case returns to $\eta\propto M^{-2}$. The simulation of
Fig.~\ref{fig:sim3}b confirms the separation: the differential cell tracks
$M^{-1}$ while the best-case photorefractive medium (optimal equalized exposure
schedule) tracks $M^{-2}$, a gap growing as $M$.

Because the coded layer applies the recall softmax globally (Sec.~\ref{sec:arch}),
the code phases are free, and they directly set the peak-to-average ratio of the
recorded superposition. A low-peak (Schroeder-type) phase assignment pulls even
the hard-clipping case back onto the ideal $M^{-1}$ slope (Fig.~\ref{fig:sim3}b),
so the code design of Sec.~\ref{sec:arch} and the recording physics here are not in
tension: the distributed codes that the coded layer reads without difficulty are
exactly those that minimize the recording dynamic-range cost. This $M^{-1}$
scaling was reached once before by digital voxel-printing that pre-computes the
final index pattern and writes it offline~\cite{Dinc2024}; the contribution here
is to recover it \emph{in situ} by an analog mechanism acting in the recording
step itself.

A recording cell performs four functions in series. \emph{Reception}: receptors at
engineered positions sample the recording field—the locus of the nonlocality of
Sec.~\ref{ssec:gradient}, since the gradient demands receptors at finite separation.
\emph{Combination}: a summing junction forms the signed, zero-sum weighted sum of
the receptor signals, realizing the gradient kernel and the DC null.
\emph{Storage}: the combined signal drives a state variable that integrates the
exposure history according to Eq.~\eqref{eq:active}, with $\tau$ set per cell.
\emph{Modulation}: that state variable modulates an optical property seen by the
read-out beam, setting the diffraction efficiency. We consider physical cell designs to achieve these functions in the next section. 

\section{Physical realization: the opposing-diode cell}
\label{sec:cells}

Section~\ref{sec:medium} traced the photorefractive $M^{-2}$ falloff to a single
fact: storage is carried by mobile photo-charge, and the same carriers that build the
space-charge field also conduct it away. The decay rate $bI$ is a photoconductive
(dielectric) relaxation $\sigma/\varepsilon$ whose conductivity $\sigma\propto I$ is
set by the very illumination that is performing the reading. Decay is locked to the read beam
because storage and leakage share one carrier population and one transport pathway;
the two pathologies of Sec.~\ref{sec:medium}—read--erases-write and shared-reservoir
depletion—are two faces of this one coupling. To prevent relaxation of the charge distribution under constant illumination, one must stop photogenerated charge from migrating back to its origin. This can be done by building the recording material around 
\emph{opposing-diode cells}, {\em i.e.}, a pair of rectifying junctions of opposite polarity
driving a shared store. 

We introduce a simple proof-of-concept ``macroscopic'' unit cell that decouples write from decay and makes the memory state immune to common-mode illumination erasure. We monitor the voltage across a storage capacitor as a function of time for different illumination scenarios to demonstrate three of the four key functionalities of each cell that were described in the previous section: reception, combination, and storage. The fourth and final function, modulation, involves driving some voltage-dependent index modulation material. Electro-optic modulation is well-characterized in spatial light modulators. Here, we neglect this function and instead demonstrate three novel capabilities that are critical for the memory cell. We assume that the index modulation of whatever effect is utilized (liquid crystal, electro-optic, or micromechanical) is proportional to the driving voltage provided by our circuit.

\begin{figure}
    \centering
    \includegraphics[width=0.5\linewidth]{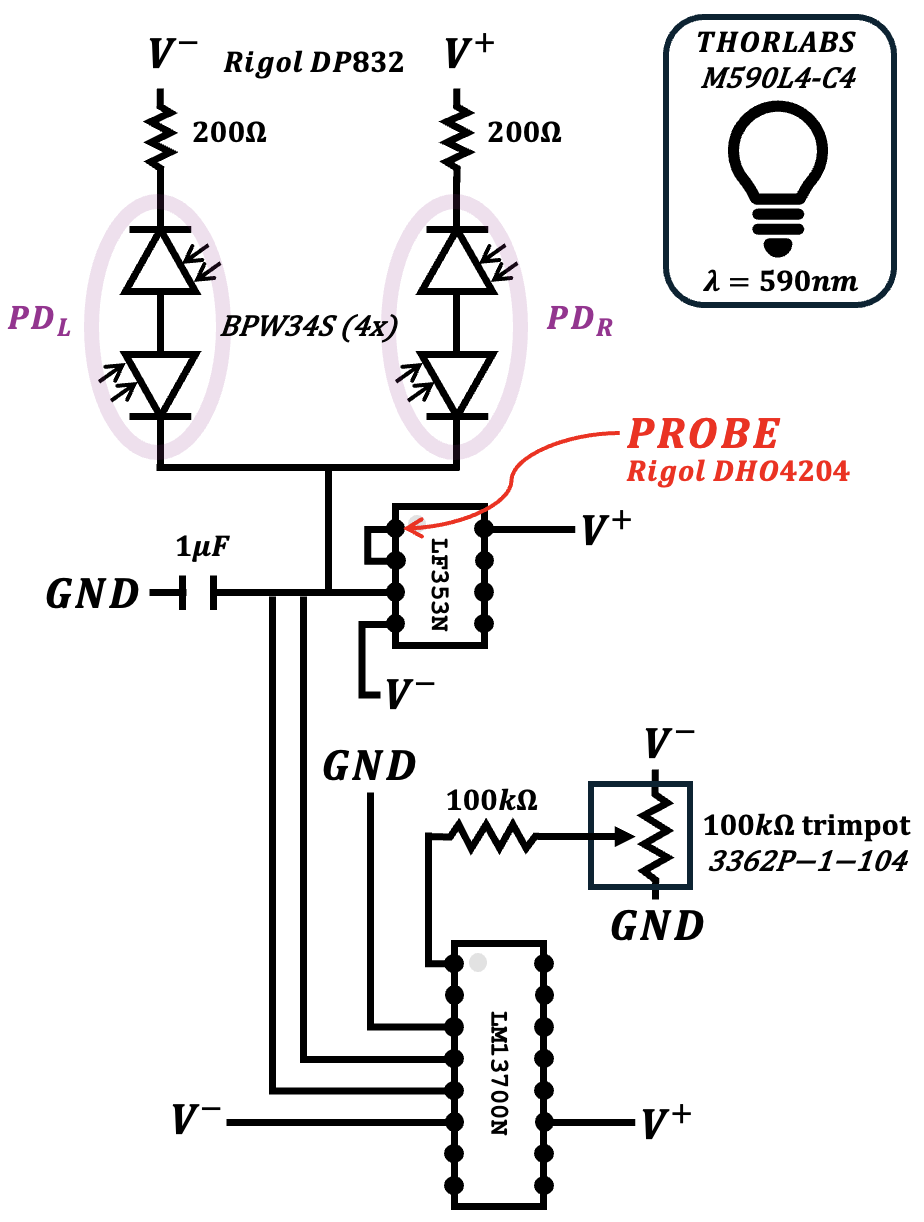}
    \caption{Schematic of our experimental setup which includes the implemented circuit and our illumination source. The source illuminates the circuit from above. }
    \label{fig:exp-setup}
\end{figure}

As we have discussed, our circuit should consist of two separate photodetection sites, a storage mechanism that remembers the state variable, and a means by which that state variable accumulates and decays over time. Critically, the state variable should respond to the \textit{spatial derivative} of the intensity across the two photodetection sites, thereby rejecting any DC components and protecting the current state from self-erasure. To do this, we build the circuit shown in Fig. \ref{fig:exp-setup}.
Two pairs of back-to-back photodetectors (BPW343S) act as the two detection sites. These two distinct sites are labeled for reference as $PD_L$ and $PD_R$ for the left and right groups and are marked in purple in the schematic. Because they are wired as back-to-back series pairs, their internal dark currents are structurally opposed. Any spontaneous leakage current from one diode is blocked by its reverse-oriented partner. This forces the net parasitic leakage of the sensor array to near-zero. These sites are wired in a ``push-pull'' configuration with a common node that is shared by the capacitor. This capacitor acts as the memory storage mechanism. The write stage is isolated from the decay stage by the OP-AMP buffer (LF353N). The decay mechanism is independently controlled via a voltage-controlled current source (VCCS) and is tuned with a trimmer potentiometer (trimpot, 3362P-1-104). An operational transconductance amplifier (OTA, LM13700N) is used as this VCCS. We provide power to the circuit with a programmable DC power supply (Rigol DP832). We monitor the voltage of the circuit over time with an oscilloscope probe (Rigol DHO4204). We evenly illuminate the photodetector sites from above with a $\lambda = 590nm$ LED (Thorlabs M590L4-C4). In summary, we have a write path that is proportional to the spatial derivative of the illumination across the two groups, and a decay path that is independently controlled with a customizable leak. If the two groups see the same power, current flows into and out of the common node at a constant rate, and no charge buildup occurs across the capacitor. Any inequality in this common-node junction will lead to charge buildup (which is our state variable memory). At the same time, charge is being leaked from the common node with the VCCS, and we can independently control this leak to make the charge persist for as long or short as we like with the trimpot. Importantly, once a memory has been written from an illumination imbalance, any DC term across the two sites will not erase that state because the decay is now independently controlled by the leak. To demonstrate this, we monitor the voltage over time of the circuit under different illumination conditions. 

\begin{figure}
    \centering
    \includegraphics[width=1\linewidth]{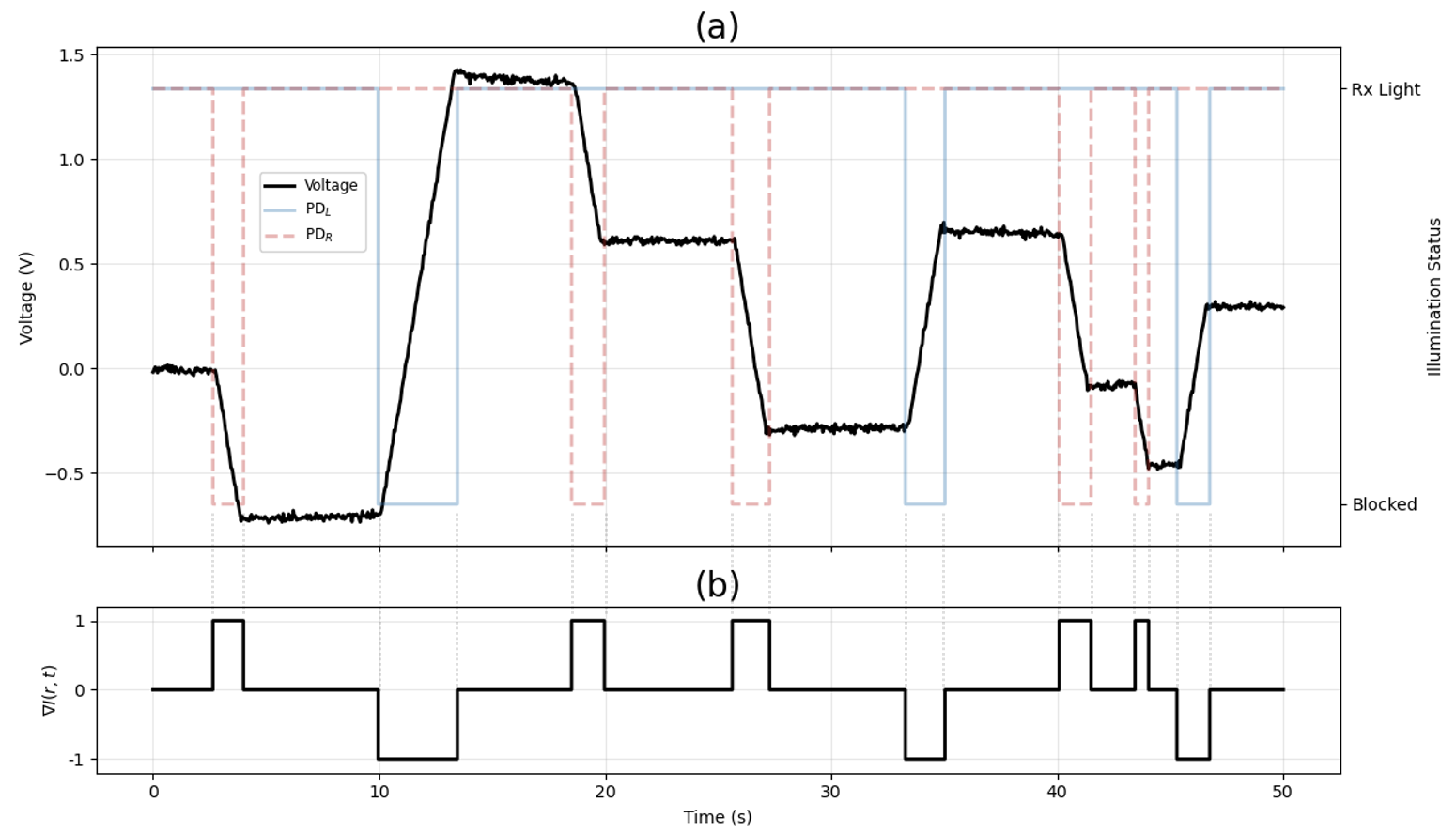}
    \caption{Tracking the circuit voltage over time with the corresponding illumination states of each photodetector group. (a) Capacitor voltage as a function of time. b) The spatial derivative of the intensity across the two sites as a function of time, $\nabla I(\textbf{r},t)$. }
    \label{fig:exp-results-1}
\end{figure}

To illustrate how the voltage over time behaves for our circuit, we choose a binary illumination modality: either light is being received by one of the PD groups, or it is being blocked. Of course, the sites can accept a continuous range of intensities, but we restrict ourselves to this simple situation for demonstration purposes. We show these results in Fig.~\ref{fig:exp-results-1}. Fig.~\ref{fig:exp-results-1}(a) shows the capacitor voltage as a function of time as the intensity on the photodetector groups varies. Superimposed on this plot on the opposite y-axis are the illumination states of the photodetector groups. The group is either receiving light (Rx Light) or being blocked. The decay time is set to be extremely long, making the state effectively frozen where it lands with respect to this time interval. When $PD_L$ is receiving light, but $PD_R$ is blocked, there is a decay in voltage. When $PD_R$ is receiving light, but $PD_L$ is blocked, there is a rise in voltage. When both are receiving light, there is no change in system voltage and the circuit holds and remembers that state until there is a difference in irradiance levels between the two groups. Correspondingly, we plot $\nabla I(\textbf{r},t)$ in Fig.~\ref{fig:exp-results-1}(b) for this specific illumination scenario. When $\nabla I(\textbf{r},t) > 0$ there is a decay, when $\nabla I(\textbf{r},t) < 0$ there is a rise, and when $\nabla I(\textbf{r},t) = 0$ there is no change. Transitory events are marked across the two plots with faint blacked dotted lines.

\begin{figure}[]
    \centering
    \includegraphics[width=1\linewidth]{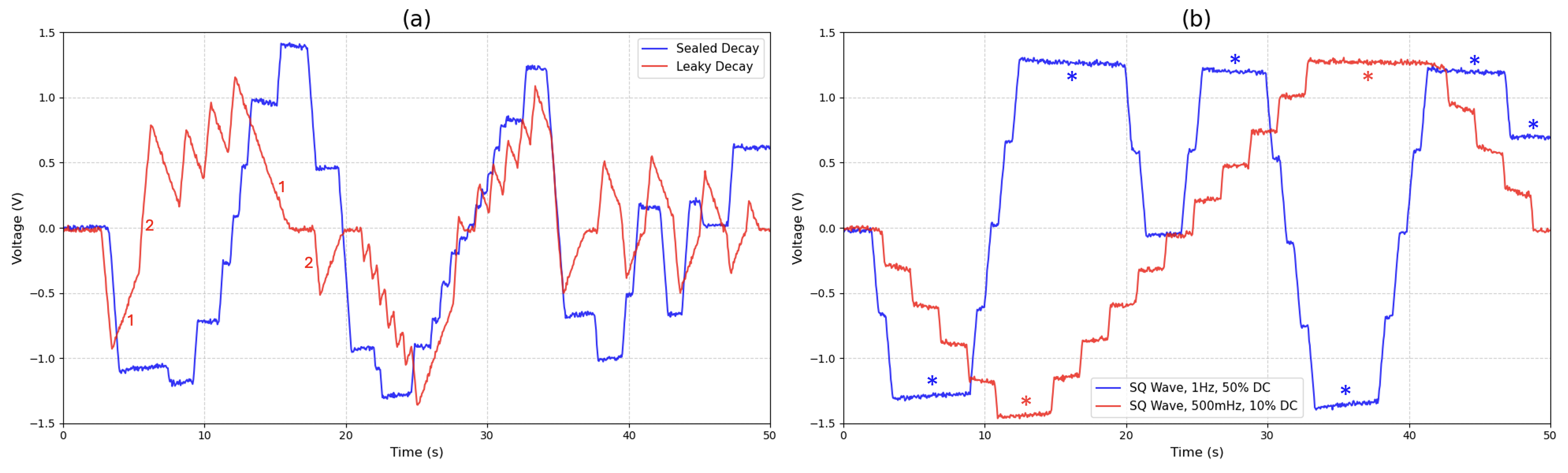}
    \caption{(a) Experimental results for two different circuit conditions with binary modulation.  (b) Experimental results after connecting a function generator to the light source and observing the voltage trend of the circuit for different pulse settings. }
    \label{fig:exp-results-2}
\end{figure}

We can also make the state decay faster by controlling the resistance of the trimpot shown in Fig. \ref{fig:exp-setup}. We show results that compare two different decay settings of the circuit in Fig.~\ref{fig:exp-results-2}(a). The “Sealed Decay” dataset was taken with the trimpot turned to knob setting 10 (maximum). The “Leaky Decay” dataset was taken with the trimpot turned to knob setting 7.5. The leaky dataset decays toward the baseline 0V value after each write event, which is undesirable for long-term memory. For each separate dataset, we arbitrarily insert and remove an opaque blocker over the two PD groups. We observe that the decay trend of the sealed dataset demonstrates the memory behavior we want: a change in the memory that corresponds to a spatial derivative of the incident field and a response that does not erase with a constant background. The light was on for the entirety of each experiment. Covering the $PD_L$ photodetector group leads to a rise in voltage and covering the $PD_R$ group leads to a decay. Concerning the leaky decay dataset: there are two unique kinds of transitory events. The first is a voltage change induced primarily by the natural decay of the circuit under common illumination. The second is a voltage change induced by a difference in received illumination. Here, the former has a slower change than that of the latter. Of course, these two effects are happening simultaneously. To illustrate this, we annotated some of these events with 1: leaky passive transition and 2: driven transition. The power of the light source was kept the same for both the sealed and leaky experiments.
 
Fig.~\ref{fig:exp-results-2}(b) shows results with dynamic modulation of the light source intensity driven by an external function generator. In the blue curve, we drive the source with a square (SQ) wave that has a frequency of 1Hz and a duty cycle (DC) of 50\%. In the red curve, the source is driven with a SQ wave with a frequency of 500mHz and a DC of 10\%. Any time the voltage goes up, that means the $PD_L$ group was covered while the light was pulsing, and any time the voltage drops it means the $PD_R$ group was covered during a light pulse event. When neither is covered, such as in ``flatline'' events marked by the asterisk ($\ast$), the memory is retained under constant illumination: the illumination continues to pulse at the prescribed rate, but there is no state change because there is no difference in received intensity (neither of the PD groups are being covered). The rise and fall events of the blue curve are longer because of the larger dose received during each ``on'' pulse, but notice that the slopes of each rise and fall event between the two cases are nearly identical. The trimpot was set to its maximum setting (10) for the data in (b). The power of the light source was again kept the same for both of these experiments plotted in Fig.~\ref{fig:exp-results-2}(b). The irradiance measured at the photodetector plane for both experiments in Fig. \ref{fig:exp-results-2} is about $0.51-0.64\ \mu$W/mm$^2$. These values are measured with a Newport Power Meter Model 1918-R at 590nm.

To obtain these results, one must “balance” the circuit so that the constant illumination yields a 0V baseline. The circuit is very sensitive to this displacement (especially for the sealed case, because in the leaky case it just decays back to zero regardless), but it is possible to get it equalized. Then, once it is balanced, we proceeded with the data collection. To help get it equalized, it’s useful to turn the power to maximum to exaggerate any small imbalances, then reduce the power for the experiment.  The slope of each transitory voltage response when there is an illumination difference will depend on the magnitude of the intensities at each of the sites.

To review, we have demonstrated that this implemented circuit performs the desired operations we set out to achieve. \textit{Reception:} our spatially offset detection sites each respond to the light intensity at those sites. \textit{Combination:} our sites are configured to drive the state variable proportional to the spatial derivative of the intensities at each site, thereby rejecting common-mode inputs. \textit{Storage:} our storage mechanism has a decoupled write and decay time, allowing us to make the memory as long or as short as we'd like.

While this device demonstrates the necessary memory-cell functions outlined in Sec.~\ref{sec:medium}, this macroscopic component must be somehow integrated into a functional 3D array for holographic recording. The most direct approach would be to build 2D OASLM with opposing diode cell pixels. Bragg selectivity in cascaded thin holograms is discussed by Nordin {\em et al.}~\cite{nordin1992diffraction} and Brady~\cite{BradyThesis1990}.
A single thin layer is in the Raman--Nath regime and has no angular
selectivity; selectivity is recovered across the \emph{stack}, which is the structure
factor $S(K)$ made explicit. With $N$ layers at pitch $\Delta z$, each carrying the
same association, the diffracted amplitudes add with a longitudinal phase to give
\begin{equation}
  |S(\Delta K_z)|^2
  \;=\;
  \frac{\sin^2\!\big(\tfrac12 N\,\Delta K_z\,\Delta z\big)}
       {\sin^2\!\big(\tfrac12 \Delta K_z\,\Delta z\big)},
  \label{eq:cascade-S}
\end{equation}
a main lobe at the Bragg condition $\Delta K_z=0$ of width $\sim 2\pi/L$ over total
depth $L=(N-1)\Delta z$, so angular and wavelength selectivity scale as $1/L$,
identical in form to the volume Bragg selectivity of
Sec.~\ref{sec:medium}~\cite{kogelnik1969,BradyPsaltis1992}. The cascade of thin layers
\emph{is} a volume hologram. Its one stack-specific artifact is replica (grating-lobe)
maxima at $\Delta K_z\,\Delta z = 2\pi n$; keeping them outside the $2k_0$ Ewald ball
requires $\Delta z\lesssim\lambda/2$ for full suppression, while a coarser stack
retains genuine $1/L$ selectivity over a bounded, replica-free range. 

Over the longer term, improved cell density and efficiency will be obtained by creating true volume arrays of rectified cells. 
In organic materials, a single molecule can carry a definite charge-transfer direction—the content of the
Aviram–Ratner rectifier~\cite{aviramRatner1974}, enhanced by donor/acceptor
charge-transfer-state engineering~\cite{nijhuis2022ct}. The optically driven version
is photoinduced charge separation in donor--acceptor triads, where absorption produces
a charge displacement of fixed direction~\cite{kuciauskas1998}; two such arms of
opposite polarity flanking a redox-active core form a molecular opposing-diode pair.
The metastable oxidation state of the core is the store, with $\tau$ the
back-electron-transfer time, engineerable through the
tether~\cite{bocian2011}; a photochromic moiety such as diarylethene supplies the
write gate $g(t)$~\cite{hiroyasu2020}; and the electrochromic contrast of the core is
the readout.

Alternatively, in a type-II nanocrystal the staggered band offset localizes electron and hole on
opposite sides of an interface, so the offset \emph{is} the junction---a directional
charge transfer with no metal and no hot-electron step~\cite{chemrev2024ct}. The
spatially indirect charge-separated state is the store, with $\tau$ the recombination
time set by shell thickness and barrier
height~\cite{zhu2011typeII,wu2012cdspt,hastrup2020all}; the stored charge
modulates the index through the quantum-confined Stark effect and state
filling~\cite{miller1984qcse,empedocles1997}, so the same charge that is stored sets
the diffraction efficiency. An asymmetric segmented nanorod---two arms of opposite
offset flanking a storage segment---is a single-particle opposing-diode cell, and such
rods disperse in a two-photon-polymerization resin~\cite{Sun2000}.

Operated photovoltaically, the junctions draw the energy to separate and direct charge
from the absorbed light itself, so the cell needs no per-cell bias rail. This is not a
convenience but the enabling fact for the volume: a powered circuit per voxel is
untenable at the $\sim\!10^{12}$ cells of a centimetre cube, whereas a self-powered
cell carries its own write--store--relax dynamics with no external supply. Two
boundaries keep the claim honest. First, self-powering covers the cell's internal
dynamics---write, store, and decay---but not the readout amplification: the softmax
inverse temperature $\lambda$ of Sec.~\ref{sec:arch} is the gain of the coded-layer
amplifiers, on the read side, and is not powered by the cells. Second, the
photovoltaic open-circuit voltage sets a natural per-cell ceiling on the stored state.

The directional derivative sensed by a differential cell fixes its spatial-frequency
response. For a recording fringe of grating vector $\mathbf{K}$, two receptors separated
by a baseline $\mathbf{p}_c$ and each of finite width $w$ return a differential signal with
transfer function
\begin{equation}
H_{\mathbf{p}_c}(\mathbf{K}) \;=\; 2\,\sin\!\big(\tfrac12\,\mathbf{K}\!\cdot\!\mathbf{p}_c\big)\,
\mathcal{A}(\mathbf{K}),
\label{eq:cell-transfer}
\end{equation}
the product of a baseline (gradient) factor and the single-receptor aperture form factor
$\mathcal{A}(\mathbf{K})$, which rolls off near $|\mathbf{K}|\sim 2\pi/w$. The baseline factor
nulls at $\mathbf{K}=0$—the DC rejection that motivates the cell—peaks first at
$\mathbf{K}\!\cdot\!\mathbf{p}_c=\pi$ (a fringe of period $2|\mathbf{p}_c|$ aligned with the
baseline), vanishes for $\mathbf{K}\perp\mathbf{p}_c$, and repeats its nulls at
$\mathbf{K}\!\cdot\!\mathbf{p}_c=2\pi n$. A single fixed baseline is therefore not a uniform
high-pass but an oriented band-pass: it records only a band of grating magnitudes around
$\pi/|\mathbf{p}_c|$, oriented along $\hat{\mathbf{p}}_c$.

This is the right primitive provided the medium contains cells of diverse baseline
orientation and length whose responses jointly tile the grating-vector support the
architecture requires. The stored connections occupy the
difference set $\mathbf{K}=\mathbf{k}_{\mathrm{code}}-\mathbf{k}_{\mathrm{in}}$ fixed by the
input-cap/code-curve geometry of Sec.~\ref{sec:arch}, a compact region of the
transverse $\mathbf{K}$-plane, so the baselines need cover
only the transverse extent. Coverage is then a filter-bank construction in two parts.
Radially, an octave ladder of baselines $|\mathbf{p}_m|=2^{-m}p_0$ places its transmission
peaks $|\mathbf{K}|=\pi/|\mathbf{p}_m|$ at a geometric sequence of spatial frequencies; because
each baseline's first null at $2\pi/|\mathbf{p}_m|$ falls on the next-shorter baseline's peak,
the union of $|H|$ has no radial gaps, and $\sim\log_2(K_{\max}/K_{\min})$ octaves span from
the coarsest fringe (set by the signal aperture) to the finest (set by the data-plane
bandwidth, below the $2k_0$ Ewald limit). Azimuthally, the $\cos\psi$ dependence of the
baseline factor for $\mathbf{K}$ at angle $\psi$ to $\mathbf{p}_c$ means orientations spaced by
$\Delta\theta\lesssim 30^\circ$ keep every grating direction within a strongly responding
population. The medium thus carries of order $N_\theta N_{\mathrm{oct}}$ distinct cell
types-a few tens-interleaved throughout the volume; with $\sim10^{12}$ cells available,
each (orientation, scale) band is populated at high redundancy, the multiplexing cost being
only the constant $1/(N_\theta N_{\mathrm{oct}})$ share of the cell budget devoted to each
band. Concretely, the highest-frequency bands set the most demanding dimensions: reaching transverse
gratings of magnitude $|\mathbf{K}|_{\max}$ calls for baselines $|\mathbf{p}_c|\sim\pi/|\mathbf{K}|_{\max}$
and receptor widths $w\lesssim 2\pi/|\mathbf{K}|_{\max}$, which at a silicon-detector-compatible
visible wavelength $\lambda_0\approx0.6~\mu\mathrm{m}$ in a host of index
$n_0\approx1.5$ fall, as $|\mathbf{K}|_{\max}$ approaches the Ewald limit $2k_0$, to
$|\mathbf{p}_c|\sim\lambda_0/4n_0\approx0.1~\mu\mathrm{m}$ and
$w\lesssim\lambda_0/2n_0\approx0.2~\mu\mathrm{m}$-both well subwavelength-while the coarser
bands up the octave ladder use proportionally larger cells of order a micron; this is consistent
with, rather than contrary to, the medium's design, since a weakly absorbing volume that mostly
modulates rather than absorbs the beam requires receptors that intercept only a small fraction of
the recording flux, and such receptors are naturally subwavelength in extent.

Because the cell is self-powered, its energy cost is set by the charge needed to establish the
store. Writing a state of $n_e$ stored elementary charges (equivalently $Q=C_c V$ for a cell of
capacitance $C_c$ at voltage $V$) photovoltaically at quantum efficiency $\eta$ requires
$\sim n_e/\eta$ absorbed photons, an energy
\begin{equation}
E_{\mathrm{cell}} \;\sim\; \frac{n_e}{\eta}\,\frac{hc}{\lambda}.
\label{eq:cell-energy}
\end{equation}
A molecular or nanocrystal store of $n_e\sim10^2$ charges costs tens of attojoules; a
capacitive cell, $C_c\sim1$~fF at $V\sim0.5$~V ($n_e=C_cV/e\sim10^3$--$10^4$), costs of order
femtojoules-both within the attojoule-femtojoule envelope of efficient
optoelectronics~\cite{miller2017attojoule}. The whole volume therefore writes for
$E_{\mathrm{vol}}=N_{\mathrm{cell}}E_{\mathrm{cell}}\sim10^{12}\times(10^{-17}\text{--}10^{-15}~\mathrm{J})
\sim 10~\mu\mathrm{J}$--$1$~mJ per complete rewrite; at a write time $t_w$ the recording beam
need only deliver $E_{\mathrm{vol}}/t_w$ to the volume---milliwatts for second-scale writing,
watts for millisecond writing. The only standing draw is the leak that enforces the forgetting
time, $P_{\mathrm{leak}}\sim C_cV^2/\tau$ per cell, $\sim10^{-16}$~W for a femtofarad cell at
$\tau\sim1$~s, hence sub-milliwatt summed over $10^{12}$ cells and---because the decay is
decoupled from illumination---independent of the read beam. No per-cell bias rail is drawn. A
floor on the write comes from shot noise: the differential photocurrent
$m\sin(\tfrac12\mathbf{K}\!\cdot\!\mathbf{p}_c)\,i_{ph}$ for a fringe of modulation depth $m$
must exceed $\sqrt{2 e\, i_{ph} B}$, which sets a minimum photon budget per cell scaling as
$1/m^2$ and a corresponding minimum recording irradiance; the breadboard's
$\sim0.5~\mu\mathrm{W/mm^2}$ is far below what a micron-scale cell would use, so a buildable
volume trades higher irradiance for a shorter exposure at fixed per-cell photon count. The read
side---probe beam and the coded-layer softmax amplifiers of gain $\lambda$---is externally
powered and lies outside this budget, consistent with the self-powering boundary noted above.

These three routes---the discrete cell demonstrated above, a cascade of
opposing-diode OASLM layers, and a true volume of molecular or nanocrystalline
rectifiers---form a ladder of decreasing maturity and increasing density. The
discrete cell fixes the physics: reception at a finite baseline, zero-sum
combination with a DC null, and storage on a clock decoupled from the read
beam. The OASLM stack inherits these functions unchanged and recovers Bragg
selectivity at
a cell density set by lithography rather than by molecular dimensions; built
from established amorphous-silicon/ferroelectric-liquid-crystal
technology~\cite{Barbier1992,Landreth1992}, it is the natural vehicle for
a first system-level demonstration, at the cost of a layer count that grows with
the selectivity demanded. The volume routes promise the $\sim\!10^{12}$-cell
density that motivates the architecture, but require a recording
medium---monodisperse segmented nanorods held at known orientation, or oriented
donor--acceptor assemblies---that does not yet exist as such, even though each
constituent function has been shown in isolation.

What the discrete demonstration leaves open is the modulation function. We have
shown reception, combination, and storage in a cell whose state variable is a
capacitor voltage; converting that voltage into an index change is the one cell
function we have assumed rather than built, and it is the function on which
diffraction efficiency---and hence the dynamic range that sets the $M^{-1}$
scaling of Sec.~\ref{sec:medium}---ultimately rests. In the OASLM route the
modulator is the liquid-crystal or electro-optic layer the device already
carries; in the volume routes it is the quantum-confined Stark shift or the
electrochromic contrast of the core, co-located with the store by construction.
Closing the write--store--read loop in a single cell with the modulation stage
included---rather than the write--store--decay shown here---is the immediate
experimental priority, and it is independent of the eventual choice of volume
chemistry.

\section{Outlook}
\label{sec:roadmap}

Optical neural processing has a long and crowded history. Coherent matrix--vector engines~\cite{shen2017coherent}, diffractive networks~\cite{lin2018diffractive}, photonic-integrated
tensor cores~\cite{feldmann2021tensorcore}, reservoir computers~\cite{vandoorne2014reservoir}, and free-space attention layers~\cite{xu2023optronictransformer} have all
been proposed and, in many cases, built~\cite{fu2024optical,wetzstein2020inference}. Much of
this work pursues generality: a programmable optical fabric onto which an
arbitrary network is mapped. We have taken the opposite stance. Rather than ask
what a general optical computer should look like, we have isolated one
functional component of a neural system---associative recall---and one technical
question within it---how a volume holographic medium can store the inter-neuron
connections faithfully and densely---and pursued those to a concrete physical
mechanism.

Associative recall in its modern form is a precise
map, the dense-associative-memory retrieval step of Eq.~\eqref{eq:attention}, and we
have shown that a two-stage code-mediated cascade evaluates that map exactly,
with the inverse temperature realized as an amplifier gain. The recording problem it imposes is equally
precise---reject the self-energy, keep the connection---and a nonlocal,
gradient-responsive medium with illumination-independent decay solves it,
recovering the $M^{-1}$ efficiency scaling in situ, with a discrete-component cell demonstrating the
required physics. Each contribution is a
component-level result with a bounded claim; none of them is a complete vision
processor, and we have been explicit throughout about what is assumed---the
softmax layer, the modulation stage, a buildable volume medium---and what is
shown.

The near-term path is correspondingly concrete. First, the single-cell
demonstration of Sec.~\ref{sec:cells} should be extended through the
modulation function, closing the write--store--read loop in one cell. Second, a
small opposing-diode OASLM stack would test the Bragg selectivity and the $M^{-1}$ recording scaling against a
real superposition of shift-multiplexed pages. Third, the coded-layer
softmax---assumed here to be electro-optic---must be integrated with the two
holographic stages into a single recall path, at which point the cascade of
Sec.~\ref{sec:arch} can be measured rather than derived. The volume
chemistries are a longer program, and the OASLM stack is the bridge to them.

We believe that progress in optical neural processing is more likely to come
from this component-by-component attack---pairing a well-defined computational
primitive with the specific device physics that realizes it---than from
general-purpose optical fabrics alone. If the dense associative memory is, as we
have argued, the operation for which volume holography is the natural hardware,
then closing the device gaps identified here yields a genuinely useful optical
neural component, and with it brings practical optical neural processing
measurably closer to reality.

\section*{Data availability}

Data and simulation code used and/or analyzed during the current study are available from the corresponding author upon reasonable request.

\bibliographystyle{unsrt}
\bibliography{references}

\end{document}